\documentclass[preprint,showpacs,preprintnumbers,amsmath,amssymb,showkeys]{revtex4}
\usepackage{amsfonts}
\usepackage{graphicx, epsfig, amsmath}
\usepackage{dcolumn}
\usepackage{bm}
\usepackage{epstopdf}
\usepackage{color}
\usepackage{hyperref}
\usepackage{subfigure}
\usepackage[normalem]{ulem}
\usepackage{textcomp}

\hypersetup{
    colorlinks=true,
    linkcolor=red,
    citecolor=blue,
} 

\newcommand{\be}{\begin{equation}}
\newcommand{\ee}{\end{equation}}
\newcommand{\bs}{\begin{split}} 
\newcommand{\bea}{\begin{eqnarray}}
\newcommand{\eea}{\end{eqnarray}}

\newcommand{\bo}{\raise-1mm\hbox{\Large$\Box$}}

\newcommand{\f}[2]{\frac{#1}{#2}}

\newcommand{\bd}{\boldsymbol}

\newcommand{\eqn}[1]{(\ref{#1})}

\begin{document}



\title{On a Nonlinear Newtonian Gravity and Charging a Black Hole}

\author{Michael R.R. Good}
	\email{michael.good@nu.edu.kz}
	\affiliation{Department of Physics, School of Science and Technology,\\ Nazarbayev University, Astana, Kazakhstan}

\begin{abstract}

A scalar field gravitational analog of the Reissner-Nordstrom solution is investigated.  The nonlinear Newtonian model has an upper-limit of charge for a central mass which agrees with the general relativistic condition required for the existence of the black hole horizon. The maximum limit for accumulation by bombardment of charged particles is found.  The aim is to investigate the resulting physics after severing the effects of curvature from the effects of energy-mass equivalence.


\end{abstract}
\maketitle

\section{Introduction}

Newtonian gravity is linear and completely described by a single 
scalar, $\phi_N$, whose only source is mass.
Spacetime curvature as described by general relativity is nonlinear, tensorial, and originates from a variety of sources: mass, fields, angular momentum, etc. 

It was shown by Peters \cite{peters!} that extending the basic notion of the source of Newtonian gravity to include fields as source contributions results in a nonlinear field theory (but still one in terms of a single scalar field variable). Solutions in this non-linear Newtonian gravity (NN) representing fields in regions exterior to a highly compact mass, $M$, are found to possess singularities which have analog black hole characteristics.

It was later shown by Young \cite{young!} that if $M$ were imagined to have a net charge, $Q$, an analog black hole-like gravitational field would result in which the charge can contribute directly and significantly to that gravitational field. 

Peters' solution \cite{peters!} bears a similarity to the Schwarzschild solution found in general relativity as well as having enormous mathematical simplicity. Young's solution \cite{young!}, on the other hand, contains features exhibited by the Reissner-Nordstrom solution of the Einstein-Maxwell equations, but again with substantial mathematical simplicity offered by the elementary scalar formulation.

The motivation here is to present the implications of a simple, rational, physically justified (but ultimately flawed) analog of a general relativistic field, which is sourced by mass \textit{and} charge.  This work is not meant to extend Newtonian gravity nor general relativity (which is the unique second order theory of gravity in 3+1 dimensions \cite{lovelock}).  Rather, its purpose is to capture the utility of an investigation into the physical ramifications of an elementary model without curvature, (in the spirit of, for example, the self-consistency investigation performed by Franklin \cite{Franklin:2014ora} or the simple inclusion of energy-mass relationship to derive the Lorentz factor as was done by Cross \cite{cross}).  

The gravitational field of a charged analog black hole leads to new issues in the dynamics of charged particles interacting with the field. For example, a particle of mass $m$ and charge $q$ will experience the competing effects of gravitational `field-mass'-attraction and Coulomb charge-repulsion.  The new dynamics are reminiscent of the Reissner-Nordstrom solution in the weak-field limit which can be attractive or repulsive, i.e. the $M/r$ and  $Q^2/r^2$ pieces have opposite sign.

Consider starting with an electrically neutral black hole and 
gathering charged particles, the resulting charge build-up within the black hole would have the effect of eventually repelling additional charges. If this were not so, the central mass could ultimately become ``over-charged'' and consequently destroyed, well-known to be an impossibility in general relativity \cite{wald1977} \cite{veronica} \cite{isoyama}. 

We outline how this note will be directed toward determining the effects of energy-mass equivalence on a maximum charge that accretion can possibly add to an analog black hole. Sec.~(\ref{sec:II}) is mostly a pedagogical survey and set-up to review but also to verify the nonlinear Newtonian theory. Introducing the most basic solution and method in Sec.~(\ref{sec:IIA}), standard SI units are adopted. In Sec.~(\ref{sec:III}) and Sec.~(\ref{sec:IV}), dimensionless quantities and natural units are embraced where we find the low energy and high energy charge bombardment limits, respectively.  Sec.~(\ref{sec:V}) compares the low energy case to the high energy case.  Sec.~(\ref{sec:VI}) briefly contrasts the solution with an alternate scalar gravity, while Sec.~(\ref{sec:VII}) derives and compares the general relativistic result with the nonlinear Newtonian result. 


\section{The Nonlinear Newtonian Formulism} \label{sec:II}
\subsection{Peters' Potential}\label{sec:IIA}
The Newtonian gravitational field due to localized mass distribution $\rho 
_{m }$ is completely described by the potential function, $\phi_{N }$, 
satisfying the linear Poisson equation,
\be
\nabla^{2}\phi_{N }= 4\pi G \rho _{m} \label{poisson}.
\ee
Exterior to spherically symmetric $\rho _{m}$ representing total mass M, this has 
solution,
\be
\phi_{N}(r) = -\f{G M}{r}.
\ee
This potential is independent of charge and angular momentum.  The sole source of Newtonian gravity is mass.  

Let us include the effective mass of the field (produced by and surrounding the mass itself).  The energy density stored 
in the gravitational field will be given by \cite{peters!},
\be
u = -\f{1}{8\pi G}(\nabla \phi_{N})^{2}.  
\ee
Taken in conjunction with the energy-mass relationship, $E=m c^2$, the energy density can be interpreted as a field-mass density,
\be
\rho _{f } = \frac{u}{c^2} = -\f{1}{8\pi G c^2}(\nabla \phi_{N})^{2}. \label{field1}
\ee
Allowing the mass density, the source of Newtonian gravity, to include that associated with the field ($\rho _m \rightarrow \rho _m + \rho_f$), Eq.~\eqn{poisson} and Eq.~\eqn{field1}  combine to yield the nonlinear field equation,
\be
\nabla^{2}\phi + \f{1}{2c^2}(\nabla \phi)^{2} = 4\pi G \rho _{m}.\label{nfe} 
\ee
This has a spherically symmetric solution, exterior to the spherical region of $M$, ($\rho _{m} = 0$),
\be
\phi(r,M) = 2c^2\ln\left(1-\f{G M}{2c^2 r}\right). \label{peterspotential} 
\ee
The Peters potential, Eq.~\eqn{peterspotential}, 
$\phi(r)$ $\to$ $\phi_{N}(r) = -GM/r$ in the weak gravity limit $M \to 0$.  The same holds true for the large $r\to \infty$ limit. 

Interestingly, a body of compact mass $M$ having radius $R < GM/(2c^2)$, would be confined to an infinitely deep potential well. Therefore this NN gravity leads in a simple way, to the existence of an analog-type black hole horizon which is somewhat similar to the Schwarzschild black hole horizon of general relativity, (but completely absent from Newtonian gravity). 

We stress the likeness of the two singularities is, of course, ultimately superficial. While the Schwarzschild metric has a singularity at the horizon, it is a coordinate singularity that is easily removed with a suitable change of coordinates: the potential at the horizon of a Schwarzschild black hole is not infinite. 

Furthermore, in the Schwarzschild geometry a description is possible inside the compact mass. In this scalar case the gravitational potential becomes infinitely negative at the horizon, but the interior has no real solution, so it is not clear what motion the particles (and charges) falling under the horizon will assume. This asymmetry weakens the meaning of the apparent analog of the solution.  

In addition, the self-sourcing of Eq.~(\ref{field1}) that results in Eq.~({\ref{nfe}) is, in a certain sense, inconsistent with itself, as has been shown by Franklin \cite{Franklin:2014ora}.  Nevertheless, we are interested in investigating this toy model of a black hole due to the ability to separate the non-linearity of self-coupling from the effects of curved spacetime with a simple scalar field.

\subsection{Young's Potential}
The NN gravity counterpart to the Reissner-Nordstrom (RN) black hole \cite{young!} is realized by including the electric field energy density \cite{jackson},
\be
u_{e}=\f{1}{8\pi}(\nabla \Phi )^{2}, \label{field2}
\ee
where $\Phi = Q/r $ is the electrostatic scalar potential. We have expressed this in natural units $G=c=4\pi \epsilon_0=1$, for clarity because the exterior ($\rho_m = 0$) field equation, Eq.~(\ref{nfe}) is now nicely expressed as,
\be
\nabla ^{2}\phi + \f{1}{2}(\nabla \phi)^2 - \f{1}{2}(\nabla \Phi )^2 = 0.
\ee
This has a spherically symmetric solution, representing the 
gravitational field of a mass $M$ carrying uniform charge $Q$, 
\be
\phi(r,M,Q) = 2\ln\left[\cosh\left(\f{ Q}{2r}\right)-\f{M}{Q}\sinh\left(\f{Q}{2r}\right)\right]. \label{youngpotential}
\ee
The potential given in Eq.~\eqn{youngpotential} is insensitive to the sign of the charge and its limiting form is identical to the Peters potential, Eq.~\eqn{peterspotential},  as $Q \rightarrow 0$. The argument of the logarithm is also identical to the potential found by Franklin \cite{Franklin:2014ora}. 

The same Peters-type analog black hole behavior occurs because $\phi$ exhibits a singularity as the argument of the log becomes zero. This is the case, 
for given $Q$ and $M$, at $r=R_{BH}$, where 
\be
R_{BH} \equiv  Q /\ln\left(\f{M +  Q}{M- Q}\right). \label{RBH}
\ee
In order that such a radius exist, the argument of the log in Eq.~\eqn{RBH} must be positive. This condition requires that such a radius can exist in NN gravity only if the mass is not over-charged. Expressing this condition as the inequality
\be
M^2 > Q^2, \label{chargemassratio}
\ee
reveals it agrees with the condition found in the 
Reissner-Nordstrom solution \cite{Misner:1974qy}. Should the condition be violated in the RN solution, then no black hole 
exists. 

For perspective, in SI units of Coulombs per kilogram, this condition is
\be \frac{Q^2}{M^2} < 4\pi\varepsilon_0 G \approx 10^{-20} [\textrm{C/kg}]^2. \ee
In the case of a proton, for example, the squared charge-to-mass ratio is 
\be \frac{q^2}{m^2} \approx 10^{16}[\textrm{C/kg}]^2, \ee
grossly violating the RN restriction.  Thus, as is well-known, a proton is not a black hole, nor is any other elementary charged particle: $q/m \gg Q/M$. 

Moreover, Eq.~(\ref{RBH}) can also become smaller than classical point charge radius, $R_c \sim Q^2/M$ (see Levine \cite{Levine}), while a RN black hole does not, (see Hod \cite{Hod}).  The charge-mass ratio when $R_c = R_\textrm{BH}$ occurs is $Q/M \tilde{0.648}$.  This is good reason to be suspect if this analog black hole becomes anywhere near this saturated with charge. 

We define the dimensionless variables, 
\be
\lambda \equiv \f{r}{R_{\textrm{NN}}}, \qquad \sigma  \equiv \f{Q}{M}, \qquad \Gamma \equiv \frac{q}{m}, \label{rholambda}
\ee 
where $R_{\textrm{NN}} \equiv M/2$ (i.e. the uncharged NN characteristic radius (which, for note, is also the general relativistic spherical black hole radius measured in isotropic coordinates \cite{peters!}- where angles in constant time hyper-slices are represented without distortion \cite{Misner:1974qy}). The potentials of masses with and without charge can then be written, respectively, as
\be
 \phi(\lambda,\sigma) = 2 \ln\left[\cosh\left(\f{\sigma}{\lambda}\right)-\f{1}{\sigma}\sinh\left(\f{\sigma}{\lambda}\right)\right], \label{youngPOT}
\ee
and
\be \phi(\lambda) = 2 \ln\left(1 - \f{1}{\lambda}\right)\label{peterPOT}. \ee
They are plotted in Fig.~(\ref{POTS}). The inequality given in Eq. \eqn{chargemassratio} restricts $\sigma$ values to the range $0 \le \sigma < 1$, with the upper limit corresponding to a highly-charged central mass. 

\begin{figure}[h]
\centering
\includegraphics[width=3.2in]{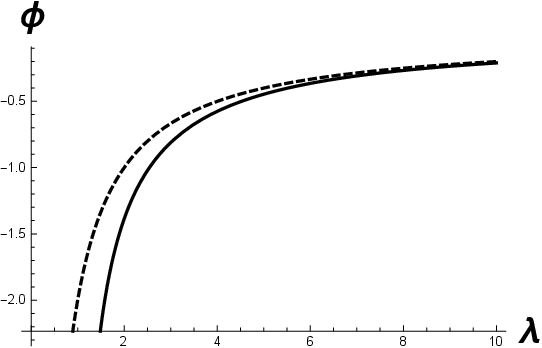}
\caption{ The potentials Eq.~(\ref{youngPOT}) and Eq.~(\ref{peterPOT}) are represented by dashed and solid lines respectively, demonstrating the singularity behavior for a finite radius, near $\lambda \sim 1$. For the dashed line, $\sigma \sim 1$, corresponding to a highly charged central mass with compressed radius.  The singularities are rudimentary scalar analogs of the Reissner-Nordstrom and Schwarzschild solutions, respectively.     \label{POTS}} 
\end{figure}

\section{Non-Relativistic Particle Dynamics}\label{sec:III}

Charge accretion by a black hole will cause a buildup, and a large value of $Q$ on $M$ leads to no black hole at all. Bombardment by $N$ charges, each having charge $q$ and mass $m$, increases the charge, $Q$, and mass, $M$, by $Nq$ and $Nm$, respectively.  We will ignore the mass contribution, taking an initially large but zero charged mass,
\be M = M_0 \gg Nm, \qquad Q_0 = 0, \ee
and then firing in $q$'s isotropically to preserve spherical symmetry. The value of $Q$ increases as the process continues and so then will the size of the Coulomb repulsion experienced by subsequent charges.  The tiny mass addition will not alter the interaction with additional bombarding protons.

During its travel, a particle's mass $m$ will interact with gravitational potential $\phi$, and its $q$ will interact with the electrical potential, $\Phi$.

The electrical potential, $\Phi = Q/r$, is not 
altered from its elementary form by including energy density effects. 
Including energy density adds only to the total mass, altering the gravity due to the energy-mass relationship.  

Since an equivalent energy-charge relationship does not exist, the equations governing NN gravity remain coupled and asymmetric, with potential energy for any such particle as: 
\be
V(r,M,Q) = m \phi(r,M,Q) + q\Phi(r,Q). 
\ee
In the course of its motion, the particle will feel the competitive 
effects of gravitational attraction ($\phi < 0$) and Coulomb repulsion ($q Q > 0$).  

Since the fields are functions of $r$ only, the particle's total energy,
\be E = T+V, \label{TE}\ee
will be conserved, where physically realistic motion can occur only when the kinetic energy $T\geq 0$, and $T$ decreases to zero along $r$ until the critical distance, $r_{c}$, giving,
\be
E = V(r_{c},M,Q). \label{turnard}
\ee
In the case of a particle approaching $M$ from $r=\infty$, where $V = 0$, we may also take the initial kinetic energy, $T_0 = 0$, or at least negligible relative to the rest-mass for a slow-moving, `non-relativistic' particle, $T_0 \ll m$. Therefore, the initial energy, $E =T_0 + m$, will be,
\be E = m, \label{Eism}\ee
and chosen as the appropriate maximum for `non-relativistic initial speeds' of the charged particles.   It is a maximum in the sense that in the range of $E=0$ to $E=m$ (non-relativistic non-existence of mass-energy), then $E=m$ is the maximum. Likewise, it is also associated with the `minimum' near-zero speed, or the minimum $E$ since, relativistically assuming mass-energy equivalence, $E>m$ for $T\neq 0$.  Eq.~(\ref{turnard}) corresponds to a point of turn-around; i.e. the point of closest approach, in the particles motion.

Taking any lower energy than the rest mass, $E<m$, say, e.g. the non-relativistic expression $E = mv_0^2/2$, would be ignoring the energy-mass equivalence that we have a priori assumed to make the field-energy contributions, Eq.~\eqn{field1} and Eq.~\eqn{field2}. The natural question is:

\textit{Can charges continue to be added 
indefinitely, thus destroying the black hole, or will electric repulsion eventually dominate gravitational attraction?}



For fixed mass $M$, gravitational attraction dominates for small $Q$, but as $Q$ increases, the Coulomb repulsion of subsequent charges grows in size so that reaching $R_{BH}$ becomes more difficult. 

Writing Eq. \eqn{turnard} and Eq. \eqn{Eism} explicitly, gives
\be
1 - 2\ln\left[\cosh\left(\f{\sigma}{\lambda}\right)-\f{1}{\sigma} 
\sinh\left(\f{\sigma}{\lambda}\right)\right] - 2\Gamma \f{\sigma}{\lambda} = 0. \label{youngcon}
\ee
Here $\Gamma\equiv q/m$ is constant.  The requirement that $T \ge  0$ means that an approaching charged particle will encounter a turning point in its motion at the value of $\lambda $ that causes $T$ to become zero. 


For small $Q$, $T\neq 0$, however eventually all the charges with $E = m$, will find themselves captured. As we will show, a definite limit on $Q$ exists such that no additional charges at this energy will be able to gain entry into the black hole, and $T=0$, so that Eq.~(\ref{youngcon}) holds. Eventually Coulomb repulsion will dominate and therefore it is impossible to destroy a black hole via low energy elementary charge bombardment.

\subsection{Excluding electric field-mass attraction}
Let us first neglect the gravitational attraction due to the electric field contribution and use Peter's potential rather than Young's potential. Eq.~(\ref{youngcon}) becomes

\be
1 - 2\ln\left[1-\frac{1}{\lambda}\right] - 2\Gamma\f{\sigma}{\lambda} = 0,\label{nocharge}
\ee
at the turn-around point, where $T=0$, for the case that the analog black hole cannot be destroyed.  To continue to exist, a turn-around will happen when $\sigma<1$, and $\lambda>1$, simultaneously.  

The solution of Eq.~(\ref{nocharge}) for $\sigma$ 
is plotted in Fig.~(\ref{TURN}). Eq.~\eqn{nocharge} is transcendentally separable for $\lambda$, yielding
\be \lambda = \frac{\Gamma \sigma}{\Gamma \sigma + W(-\Gamma \sigma e^{1/2-\Gamma \sigma})}, \label{lambdaofsigma}\ee
where $W$ is the product log (The $W$ Lambert function is used in recent black hole contexts, see e.g. \cite{Good:2016oey, Good:2012cp, Good:2016atu}).    The critical point occurs at the minimum ($\lambda_c,\sigma_c$), which has an analytic solution:
\be \lambda_c = \frac{W_{k}(-e^{-3/2})}{W_{k}(-e^{-3/2})+1} \approx 1.737,\label{clambda} \ee
\be \Gamma \sigma_c = - W_{k}(-e^{-3/2}) \approx 2.358, \ee
where $W_k$ is the $k^\textrm{th}$ solution product log with $k=-1$.


\begin{figure}[h]
\centering
\includegraphics[width=3.2in]{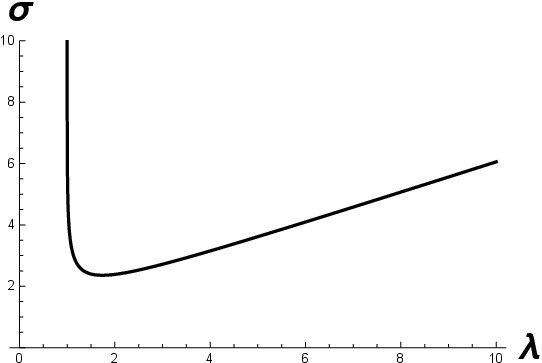}
\caption{The charge to mass ratio, $\sigma(\lambda)$, Eq.~(\ref{nocharge}), in units of $\Gamma$, for $E \sim m$.  The charges are unaffected by the charge-energy gravitational contribution, Eq.~(\ref{nocharge}), demonstrating the existence of a turn-around for $\lambda > 1$. \label{TURN}} 
\end{figure}

Therefore, closest approach occurs at about twice the radius of the NN analog black hole, while the critical black hole charge-mass ratio is also about $\sigma_c \sim 2/\Gamma$.  The result is robust, even when including the gravitational attraction from the electric field-mass, as we shall see below.

\subsection{Including electric field-mass attraction}
It is not only possible but instructive to include the charge contribution to the gravitational pull.  We assume the turnaround point, $\lambda_c$, exists, so that $T=0$, and we may use Eq.~\eqn{youngcon}.  This is solved analytically to give 
\be \lambda_c = 2 \sigma /\ln \left(\frac{(\Gamma -1) (1+\sigma)}{(\Gamma +1) (1-\sigma)}\right). \label{lambdacrit} \ee
Plugging this into Eq.~\eqn{youngcon} does not give an analytically transcendentally invertible expression for $\sigma_c$, but numerically solving for the critical charge to mass ratio gives $\Gamma \sigma_c =2.358$. 

The precise result can be accurately compared to the solution $\sigma_c$ of Eq.~(\ref{nocharge}).  We find the $\sigma_c$'s agree to $10^{-9}$ relative error.  This confirms that the gravitational attraction due to the charge contribution is negligible relative to the electromagnetic repulsion, as is unsurprising.

For perspective, in SI units, $\sqrt{4\pi\epsilon_0 G} \approx 10^{-10}  [\textrm{C}/\textrm{kg}]$, and using protons, $m/q \approx 10^{-8} [\textrm{kg}/\textrm{C}]$, the maximum charge-mass ratio of the central mass is, 
\be \sigma _c \sim \frac{2}{\Gamma} = 2 \frac{m}{q}\sqrt{4\pi\epsilon_0 G} \approx 10^{-18}.\label{low} \ee
Eq.~(\ref{low}) is a tiny number, ($\sigma_c \ll 1$), and demonstrates that `slow' initial speed charges (e.g. protons) are not capable of saturating ($\sigma_c\;  \mathtt{\sim}  1$), the central mass.   
%
\section{Relativistic Particle Dynamics}\label{sec:IV}

Consider now, depositing charge into the mass $M$ by radial bombardment of ultra high initial energy,
\be \gamma_0 \gg 1, \ee
charged particles.  For example, abundant protons in astrophysical settings have an approximate extreme maximum observed energy of $10^{20}$ eV  \cite{highenergy1} \cite{highenergy}, $\gamma_0 \sim 10^{11}$.  


Like before, a charged particle radially approaching a charged mass $M$ will experience competing gravitational and Coulomb forces, each of which is conservative and the energy inequality follows,
\be E \ge m\phi + q\Phi, \ee
or, with the initial energy $E = \gamma_0 m$,
\be
\gamma _{0} \ge \phi + \f{q}{m}\Phi.
\label{protonapproach}
\ee
\subsection{Excluding electric field-mass attraction}
We can anticipate that the gravitational attraction due to the electric field-mass will not affect the physics in the high-energy case since we have seen that even in the low-energy case it may ignored.  

Therefore consider first, only the self-energy gravitational attraction and Coulomb repulsion effects, i.e. Eq.~(\ref{nocharge}), but instead of low-energy, $E=m$, we will use high energy, $E = \gamma_0 m$, where $\gamma_0 \gg 1$:
\be \gamma_0 - 2\ln\left[1- \frac{1}{\lambda}\right] - 2\Gamma \frac{\sigma}{\lambda} = 0. \label{highenergyself}\ee
This has an analytical solution,
\be \Gamma\sigma_c = - W_{k}(-e^{-1-\gamma_0/2}), \label{nochargeGs}\ee
with $\lambda_c$ turn-around commensurate in form to Eq.~(\ref{clambda}),
\be \lambda_c = \frac{W_{k}(-e^{-1-\gamma_0/2})}{W_{k}(-e^{-1-\gamma_0/2})+1},\label{lc}\ee
where $k=-1$.  Using high energy initial particles, $\gamma_0 \gg 1$, the turn-around, Eq.~(\ref{lc}), happens right outside the characteristic radius, 
\be \lambda_c \to 1^+   \quad \textrm{for} \quad \gamma_0 \gg 1. \ee
Eq.~(\ref{nochargeGs}) becomes,
\be \Gamma \sigma_c \to \frac{\gamma_0}{2} \quad \textrm{for} \quad \gamma_0 \gg 1. \ee
This means we may negate the logarithmic term in Eq.~(\ref{highenergyself}), i.e. even the self-energy can be ignored because the high-energy $\gamma_0$ will completely swamp it.

\subsection{Including electric field-mass attraction}

Including both contributions, gravitational field-mass and electric field-mass, i.e. Eq.~\eqn{youngPOT}, we write Eq.~\eqn{protonapproach},  
\be
\gamma_0 \geq 2\ln\left[\cosh\left(\frac{\sigma}{\lambda}\right) - \frac{1}{\sigma}\sinh\left(\frac{\sigma}{\lambda}\right)\right] + 2 \Gamma \frac{\sigma}{\lambda}. \label{rescaledF}
\ee
The maximum of Eq.~\eqn{rescaledF} occurs at $\lambda \equiv \lambda_c$, Eq.~\eqn{lambdacrit}.  It is not hard, using $\Gamma \gg 1$, and, $\sigma < 1$ to show that it may be approximated by, 
\be \lambda _c \approx 1 + \frac{1}{\Gamma \sigma}, \label{lcapprox}\ee
to see that closest approach happens very close to the $\lambda \sim 1^+$, when $\Gamma \sigma \gg 1$.

While Eq.~\eqn{lcapprox} represents a good approximation of distance for closest approach to the mass $M$, it is independent of the original speed given to the charges, because it is the solution to the maximum of the right hand side of Eq.~(\ref{rescaledF}) for a certain charge-mass ratio, $\sigma$. 

Assuming $\Gamma\sigma \gg 1$, the turning point will occur very close to the analog black hole, $\lambda \approx 1^+$, from Eq.~\eqn{lcapprox}.  Therefore the maximum value of the right hand side in Eq.~(\ref{rescaledF}) is
\be
\gamma_0 \approx 2\ln \frac{\sigma ^{2}}{3} + 
2\Gamma\sigma.
\ee
An upper limit exists because the RHS is not limited by $\sigma = 1$ but by $\gamma_0$. 

The maximum charge-to-mass ratio $\sigma_\textrm{NN}$ is found by recognizing the natural log will be dominated by $2\Gamma\sigma _{\textrm{NN}}$ so that
\be
\gamma_0 \approx  2\Gamma\sigma _{\textrm{NN}}. 
\ee
Therefore, as $\Gamma \equiv q/m$,
\be \sigma_{\textrm{NN}} = \frac{Q}{M} = \frac{\gamma_0 m }{2q}. \label{NN}\ee
For perspective, we explicitly plug in $\gamma_0 \sim 10^{11}$ and use SI units: 
\be \sigma _{\textrm{NN}} = \frac{\gamma_0 m }{2q}\sqrt{4\pi \epsilon_0 G} \approx 10^{-8}.\label{high} \ee
The analog black hole will stop all extreme ultra-high energy cosmic ray protons, that is, its critical charge-to-mass ratio is extremely small, $\sigma_\textrm{NN} \ll 1$.  

\section{Comparing Low-Energy With High-Energy}\label{sec:V}
Despite the maximum limits being so tiny, it is helpful to contrast the upper limit of charge for the low-energy protons Eq.~\eqn{low} with that of the high-energy protons Eq.~\eqn{high}, which differ by $\sim 10$ orders of magnitude. 
%

%
It follows that with decreasing energy, 
the less capable the protons are of overcoming 
the Coulomb force. The less energetic protons are unable to saturate the black hole as fully as the highly energetic protons, (i.e. drenching vs dribbling). This explains the reason for a much smaller charge to mass ratio for low-energy protons.

The essential contrasting assumptions for low-energy and high-energy were $E \sim m$ versus $ E \sim \gamma_0 m$, respectively, along with the recognition that for $\gamma_0 \gg 1$ a turn-around point occurs near $\lambda \sim 1^+$ (in contrast to $\lambda \sim 2$ for $E\sim m$).  

Interestingly, the high energy case is physically simplified because the gravitational self-field and charge-field contributions are negligible compared to the Coulomb repulsion. 

In the low energy case, however, only the electric field-mass contribution to the attraction is negligible relative to Coulomb repulsion.  The gravitational field-mass contribution to attraction plays a crucial role in determining the turn-around point in the case of $E\sim m$.  

The differences from inclusion of contributions is physically note worthy: both are negligible in the high-energy case, while the self-field contribution is crucial in the low-energy case. This leads to different turn-around points for the different energies and justifies that in the limit $\gamma_0 \to 1$, the high-energy ratio does not equal the low-energy ratio, $\gamma_0 /(2\Gamma) \neq 2/\Gamma$.
\section{Alternate Self-Sourced Scalar}\label{sec:VI}
A comparative element can be made with an alternative non-linear scalar solution \cite{Franklin:2014ora}.  This solution was first considered by Einstein \cite{AE} and re-derived in \cite{Freund:1969hh}. The field equation has its own energy density as a source, obtained from a Lagrangian.  We write the potentials with charge: 
\be \varphi(\lambda,\sigma) = \left[\cosh\frac{\sigma}{\lambda} - \frac{1}{\sigma}\sinh\frac{\sigma}{\lambda}\right]^2,\label{franklincharged} \ee
and without charge,
\be \varphi(\lambda) = 1- \frac{2}{\lambda} + \frac{1}{\lambda^2},\label{franklinuncharged}\ee
respectively in our dimensionless variables.
Here the uncharged potential, Eq.~(\ref{franklinuncharged}), has a finite minimum at $\lambda =1$ rather than the infinite negative potential of the NN case.  

Interestingly, the charged potential, Eq.~(\ref{franklincharged}), also has a finite minimum rather than infinite negative potential, which happens at the same finite radius, Eq.~(\ref{RBH}), consistent with the RN condition $M^2 > Q^2$. 

Like the NN solutions, these possess identical points of interest in the potentials; however, unlike the NN potentials of Eq.~(\ref{youngPOT}) and Eq.~(\ref{peterPOT}), the solutions Eq.~(\ref{franklincharged}) and Eq.~(\ref{franklinuncharged}) do not have singularities at any finite radius, see Fig.~(\ref{FRANKLIN}).  From this vantage point, the NN solutions lack self-consistency of self-sourcing and this trait is correlated with the appearance of their negative infinite potentials.

These solutions mean that a body of compact mass (e.g. $\lambda < 1$ in the uncharged case) would not be confined to a infinitely deep potential well.  Therefore, the simple bottomless trapping analog is absent in the self-consistent self-sourced case.  

\begin{figure}[h]
\centering
\includegraphics[width=3.2in]{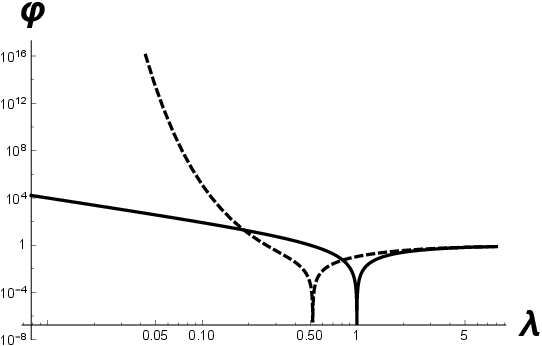}
\caption{The potentials with charge, $\sigma \sim 0.95$, (dashed) Eq.~(\ref{franklincharged}) and without charge (solid) Eq.~(\ref{franklinuncharged}), in a log-log plot demonstrating minimal finite values, interior solutions and no infinite trapping potentials at finite radius. \label{FRANKLIN}} 
\end{figure}

\section{Comparing NN with GR}\label{sec:VII}
For orientation, the Ressner-Nordstrom geometry in the weak-field limit is,
\be ds^2 \approx \left[1-\frac{2M}{x} + \frac{Q^2}{x^2}\right]dt^2-\left[1+\frac{2M}{x} -\frac{Q^2}{x^2}\right]d\bd{x}^2, \ee
using quasi-Minkowskian coordinates \cite{Sereno:2003nd}.  To lowest order in $M$ and $Q$, the signs are opposite which allow for attraction or repulsion.  It is this dynamic which has counterpart in the non-linear Newtonian theory where the electric field energy, Eq.~(\ref{field2}), acts as a mass-density, $\rho = u_e/c^2$, in the same way that the gravitational field energy was treated as a mass-density in Eq.~(\ref{field1}). 

Let us then compare the non-linear Newtonian result to the full general relativistic result. The radial 4-velocity of a particle of mass $m$, charge $q$, and energy $E$, is
\be \left(\frac{dr}{d\tau}\right)^2 = \left(E-\frac{qQ}{r}\right)^2 - \left(1-\frac{2 M}{r} + \frac{Q^2}{r^2}\right)m^2, \ee
for a charged particle falling in the Reissner-Nordstrom geometry \cite{Misner:1974qy}.  The horizon is at $1-2M/r + Q^2/r^2 = 0$, where the radial 4-velocity will also be zero.  

The condition for maximal charge build-up on the black hole occurs when
\be \frac{dr}{d\tau} = 0, \qquad  r = r_+, \ee
so that the bombardment of the last charge is unable to reach inside the event horizon.  The condition is, 
\be 
E = \frac{qQ}{r_+} = \frac{qQ}{M + \sqrt{M^2-Q^2}},
\ee
where $E$ is the energy of the charged particle at infinity, and $r_+ \equiv M+\sqrt{M^2 -Q^2}$, is the radius of the horizon.  Solving for the charge to mass ratio gives,
\be 
\sigma_\textrm{GR} \equiv \frac{Q}{M} = \frac{2 E}{q}\frac{1}{1+E^2/q^2}, 
\ee
which simplifies because $q/m \gg \gamma_0$ (i.e. even for an extreme energy cosmic ray proton, $10^{18}\gg 10^{11}$) and yields, $ \sigma_\textrm{GR} = 2E/q$, or
\be
\sigma_\textrm{GR} = 2\frac{\gamma _{0}m}{q}, 
\ee
for $E=\gamma_0 m$ high energy charges.  Comparing this with the previous high-energy NN result, Eq.~(\ref{NN}), gives
\be
\sigma _{\textrm{GR}} = 4\sigma _{\textrm{NN}}.
\ee
It would be reasonable to expect that the two descriptions would differ drastically because the foundations of the two models are so different.  The general relativistic geometric description of space-time curvature is absent in the NN description; yet the upper-limit of NN charge-mass accumulation is only four times too small.

The curvature effects present in general relativity may explain the difference. As we have seen, the non-linear Newtonian radius for an uncharged black hole $R_\textrm{NN} \equiv M/2$, is also four times smaller than the Schwarzschild radius for an uncharged black hole, $R_\textrm{GR} = 2M$:
\be R_\textrm{GR} = 4 R_\textrm{NN}. \ee
Because the charges on these black holes are far below the Reissner-Nordstrom condition, yet highly charged so that no known charges can get in, both these radii are excellent approximations for the central masses involved. 


\section{Conclusions}
Investigating physical results of any alteration to a known theoretical framework is as essential as developing the
alteration itself.  Here we calculated the maximum charge-mass ratio attainable from isotropic bombardment of charged particles onto a spherically symmetric massive source incorporating both the gravitational field-energy density and electromagnetic field-energy density in Newtonian gravity. A strict limit exists, in both the scalar and general relativistic cases (research for the case of overcharging a black hole which is very nearly saturated is ongoing \cite{wald2017}), that demonstrates the impossibility of saturation or over-charging of the central mass, even with the highest known energy charges.  

A simple merging of relativistic ideas with non-relativistic gravity yields a curious nonlinear theory of Newtonian gravity whose motivation is not without flaw, since one can and should object to using any results of special relativity in an essentially non-relativistic theory of gravitation. 

Furthermore, there is good argument that Young's extension is not the best choice of a modified theory, namely, not only are the results found in a non-valid (e.g. NN is not Lorentz covariant) relativistic field theory, there are arguably more relevant nonlinear scalar theories of gravity \cite{Franklin:2014ora}, (e.g. Sec~\ref{sec:VI}) including the prior-mentioned 1912 theory \cite{AE} which was further developed in Freund and Nambu's work on coupling to traces \cite{Freund:1969hh}. This scalar theory which also has Reissner-Nordstrom type conditions holds more self-consistency as a scalar theory of gravity than the simple Young extension. An immediate benefit of comparison underscores the availability of a scalar theory of gravity since the source is the scalar trace of the stress tensor, that leads to a full scalar field equation, as opposed to the current NN case in which the scalar field is still coupled only to the 00 component of the stress tensor.

Regardless, the usefulness of the current note is not in extending general relativity nor Newtonian gravity, but rather, in presenting a logical and physically motivating analog to a solution of general relativity; one that is simple, and whose point is to capture known `new' physics (the self-energy of the gravitational field and a charge-energy of the electric field) and reveal them in the established framework of Newtonian gravity. Addressing this general type of physically-inspirited alteration, and the evaluation of the physical contents of the resulting framework are regarded as essential elements of contemporary theoretical analysis.

While an upper limit of charge was found for the analog black hole and it was made explicitly clear that the central mass $M$ cannot be destroyed by forcing charged particles into it as described by NN, it was also demonstrated that general relativistic calculations also describe the upper limit \cite{wald2017}, and the results differ by a form factor.  

The ability to consider a simple scalar case along with a recognizable start in the Poisson equation, energy-mass equivalence, and Coulomb's law, demonstrates this is a good example of the aforementioned investigations, and may have something of a pedagogical utility.

The model in this respect is instructive because the geometric view is absent in the scalar case.  The scalar source can only be one component of the full source stress-tensor of the covariant field equations. The calculation underscores the importance of the equivalence principle as the initial motivating framework in general relativity. 

\begin{acknowledgments}
MG thanks John Young and support by the NSF, funding under Grant No. DMR-9987872, the Julian Schwinger Foundation under Grant 15-07-0000 and the ORAU and Social Policy grants at Nazarbayev University.  
\end{acknowledgments}






\begin{thebibliography}{99}

\bibitem{peters!}
  P. C. Peters, ``Where is the energy stored in a gravitational field?,''
  {\href{http://doi.org/10.1119/1.12460}{Am. J. Phys. \textbf{49}, 564-569 (1981).}}




\bibitem{young!} 
	J. H. Young, ``A charge contribution to (pseudo-) Newtonian gravity,''
    {\href{http://dx.doi.org/10.1119/1.16822}{Am. J. Phys. \textbf{59}, 565-567 (1991).}}
 
 \bibitem{lovelock} 
	D. Lovelock, ``The Four‐Dimensionality of Space and the Einstein Tensor,"
    {\href{http://dx.doi.org/10.1063/1.1666069}{J. Math Phys. \textbf{13}, 6, 874-876, (1972).}}
    
   

    
    
\bibitem{Franklin:2014ora} 
  J.~Franklin,
  ``Self-Consistent, Self-Coupled Scalar Gravity,''
  {\href{http://dx.doi.org/10.1119/1.4898585}{Am.\ J.\ Phys.\  {\bf 83}, 332 (2015).}}

\bibitem{cross}
D.~J. Cross, ``The relativistic gamma factor from Newtonian mechanics and Einstein's equivalence of mass and energy,"
  {\href{http://dx.doi.org/10.1119/1.4941828}{Am. J. Phys. \textbf{84}, 385-387 (2016).}}

  
  
\bibitem{wald1977}
R. Wald,
  ``Gedanken experiments to destroy a black hole,''
{\href{https://doi.org/10.1016/0003-4916(74)90125-0}{ Ann.\ Phys.\  {\bf 882}, 548-556 (1974).}}

\bibitem{veronica} 
  V.~E.~Hubeny,
  ``Overcharging a black hole and cosmic censorship,''
  {\href{https://doi.org/doi:10.1103/PhysRevD.59.064013}{ Phys.\ Rev.\ D {\bf 59}, 064013 (1999).}}

\bibitem{isoyama}
S.~Isoyama, N.~Sago and T.~Tanaka,
  ``Cosmic censorship in overcharging a Reissner-Nordstr\'{o}m black hole via charged particle absorption,''
   {\href{https://doi.org/ doi:10.1103/PhysRevD.84.124024}{ Phys.\ Rev.\ D {\bf 84}, 124024 (2011).}}








\bibitem{jackson} 
  J.~D.~Jackson,
  \textit{Classical Electrodynamics},
  Wiley, (1998).

\bibitem{Misner:1974qy}
  C. Misner, K. Thorne, and J. Wheeler,
  \textit{Gravitation},
Freeman, (1973).


 
 
 \bibitem{Levine}
 H. Levine, E. J. Moniz, and D. H. Sharp,
 ``Motion of extended charges in classical electrodynamics,''
  {\href{http://dx.doi.org/10.1119/1.10914}{Am. J. Phys. 45, 75 (1977).}}
 
 

\bibitem{Hod} 
  S.~Hod,
  ``Universal charge-mass relation: From black holes to atomic nuclei,''
{\href{http://dx.doi.org/10.1016/j.physletb.2010.08.044}{Phys.\ Lett.\ B {\bf 693}, 339 (2010).}}

\bibitem{Good:2016oey} 
  M.~R.~R.~Good, P.~R.~Anderson and C.~R.~Evans,
  ``Mirror Reflections of a Black Hole,''
  {\href{https://doi.org/10.1103/PhysRevD.94.065010}{Phys.\ Rev.\ D {\bf 94}, 065010 (2016).}}

\bibitem{Good:2012cp} 
  M.~R.~R.~Good, ``On Spin-Statistics and Bogoliubov Transformations in Flat Spacetime With Acceleration Conditions,''
  {\href{https://doi.org/10.1142/S0217751X13500085}{Int.\ J.\ Mod.\ Phys.\ A {\bf 28}, 1350008 (2013).}}
 
\bibitem{Good:2016atu} 
  M.~R.~R.~Good, K.~Yelshibekov and Y.~C.~Ong,
  ``On Horizonless Temperature with an Accelerating Mirror,''
  {\href{https://doi.org/10.1007/JHEP03(2017)013}{JHEP {\bf 1703}, 013 (2017).}}


\bibitem{highenergy1} 
 J. Linsley,
  ``Evidence for a Primary Cosmic-Ray Particle with Energy ${10}^{20}$ eV,''
  {\href{https://doi.org/10.1103/PhysRevLett.10.146}{ Phys.\ Rev.\ Lett.\  {\bf 10}, 4, 146, (1963).}}

\bibitem{highenergy} 
  N.~Shaham and T.~Piran,
  ``Implications of the Penetration Depth of Ultrahigh-Energy Cosmic Rays on Physics at 100 TeV,''
  {\href{https://doi.org/10.1103/PhysRevLett.110.021101}{ Phys.\ Rev.\ Lett.\  {\bf 110}, 021101 (2013).}}


 


\bibitem{AE}
A.~Einstein, 
``Zur theorie des statschen gravitationsfeldes," 
{\href{http://onlinelibrary.wiley.com/doi/10.1002/andp.19123430709/abstract;jsessionid=4C7502731D4AC2F26BD44F7087B9602A.f03t03}{Ann. Phys.
(Leipzig) 343, 443–458 (1912).}} 
The Collected Papers of Albert Einstein,
translated by Anna Beck (Princeton U.P., Princeton, 1996), Vol. 4, pp.
107-120].
  
\bibitem{Freund:1969hh} 
  P.~G.~O.~Freund and Y.~Nambu,
  ``Scalar field coupled to the trace of the energy-momentum tensor,''
  {\href{https://doi.org/doi:10.1103/PhysRev.174.1741}{ Phys.\ Rev.\  {\bf 174}, 1741 (1968).}}
  



\bibitem{Sereno:2003nd} 
  M.~Sereno,
  ``Weak field limit of Reissner-Nordstrom black hole lensing,''
  {\href{http://dx.doi.org/10.1103/PhysRevD.69.023002}{Phys.\ Rev.\ D {\bf 69}, 023002 (2004).}}




  \bibitem{wald2017} 
  J.~Sorce and R.~Wald,
  ``Gedanken Experiments to Destroy a Black Hole II: Kerr-Newman Black Holes Cannot be Over-Charged or Over-Spun,''
   {\href{https://doi.org/ doi:10.1103/PhysRevD.96.104014}{ Phys.\ Rev.\ D {\bf 96}, no. 10, 104014 (2017).}}

\end{thebibliography}
\end{document}